\documentclass{desyproc}

\usepackage[belowskip=-15pt,aboveskip=0pt]{caption}
\begin{document}
\title{IAXO - The International Axion Observatory}

\author{{\slshape J.~K. Vogel$^{1}$, F.~T. Avignone$^{2}$, G. Cantatore$^{3}$, J.~M. Carmona$^{4}$, S.~Caspi$^{5}$, S.~A. Cetin$^{6}$, F.~E. Christensen$^{7}$, A.~Dael$^{~8}$, T.~Dafni$^{4}$, M.~Davenport$^{9}$, A.~V. Derbin$^{10}$, K.~Desch$^{11}$, A.~Diago$^{4}$, A.~Dudarev$^{9}$, C.~Eleftheriadis$^{12}$, G.~Fanourakis$^{13}$, E.~Ferrer-Ribas$^{8}$, J.~Gal\'an$^{8}$, J.~A. Garc\'ia$^{4}$, J.~G. Garza$^{4}$, T.~Geralis$^{13}$, B.~Gimeno$^{14}$, I.~Giomataris$^{8}$, S.~Gninenko$^{15}$, H.~G\'omez$^{4,28}$, C.~J. Hailey$^{16}$, T.~Hiramatsu$^{17}$, D.~H.~H. Hoffmann$^{18}$, F.~J. Iguaz$^{4}$, I.~G. Irastorza$^{4}$, J.~Isern$^{19}$, J.~Jaeckel$^{20}$, K.~Jakov\v{c}i\'{c}$^{21}$, J.~Kaminski$^{11}$, M.~Kawasaki$^{22}$, M.~Kr\v{c}mar$^{21}$, C.~Krieger$^{11}$, B.~Laki\'{c}$^{21}$, A.~Lindner$^{23}$, A.~Liolios$^{12}$, G.~Luz\'on$^{4}$, I.~Ortega$^{4}$,  T.~Papaevangelou$^{8}$, M.~J. Pivovaroff$^{~1}$, G.~Raffelt$^{24}$, J.~Redondo$^{24}$, A.~Ringwald$^{23}$, S.~Russenschuck$^{9}$, J.~Ruz$^{1}$, K.~Saikawa$^{22}$, I.~Savvidis$^{12}$, T.~Sekiguchi$^{22}$, I.~Shilon$^{4,9}$, H.~Silva$^{9}$, H.~H.~J.~ten~Kate$^{9}$, A.~Tomas$^{4}$, S.~Troitsky$^{15}$,K.~van~Bibber$^{25}$, P.~Vedrine$^{8}$, J.~A. Villar$^{4}$, L.~Walckiers$^{9}$, W.~Wester$^{26}$, S.~C. Yildiz$^{6}$, K.~Zioutas$^{27}$}\\[1ex]
$^{1}$Lawrence Livermore National Laboratory (LLNL), Livermore, California, USA; vogel9@llnl.gov\\
$^{2}$Physics Department, University of South Carolina, Columbia, South Carolina, USA\\
$^{3}$Instituto Nazionale di Fisica Nucleare (INFN), Sezione di Trieste \& Universit\`a di Trieste, Italy\\
$^{4}$Laboratorio de F\'{\i}sica Nuclear y Altas Energ\'{\i}as, Universidad de Zaragoza, Zaragoza, Spain\\
$^{5}$Lawrence Berkeley National Laboratory (LBNL), Berkeley, California, USA.\\
$^{6}$Dogus University, Istanbul, Turkey\\
$^{7}$DTU Space, Technical University of Denmark, Copenhagen, Denmark\\
$^{8}$IRFU, Centre d'\'Etudes Nucl\'eaires de Saclay (CEA-Saclay), Gif-sur-Yvette, France\\
$^{9}$European Organization for Nuclear Research (CERN), Gen\`eve, Switzerland\\
$^{10}$St.Petersburg Nuclear Physics Institute, St.Petersburg, Russia\\
$^{11}$Physikalisches Institut der Universit\"{a}t Bonn, Bonn, Germany\\
$^{12}$Aristotle University of Thessaloniki, Thessaloniki, Greece\\
$^{13}$National Center for Scientific Research “Demokritos”, Athens, Greece\\
$^{14}$Instituto de Ciencias de las Materiales, Universidad de Valencia, Valencia, Spain\\
$^{15}$Institute for Nuclear Research (INR), Russian Academy of Sciences, Moscow, Russia\\
$^{16}$Columbia University Astrophysics Laboratory, New York, NY, USA\\
$^{17}$Yukawa Institute for Theoretical Physics, Kyoto University, Kyoto, Japan\\
$^{18}$Technische Universit\"{a}t Darmstadt, IKP, Darmstadt, Germany\\
$^{19}$Institut de Ci\`encies de l'Espai (CSIC-IEEC), Facultat de Ci\`encies, Bellaterra, Spain\\
$^{20}$Institut f\"ur theoretische Physik, Universit\"at Heidelberg, Heidelberg, Germany.\\
$^{21}$Rudjer Bo\v{s}kovi\'{c} Institute, Zagreb, Croatia\\
$^{22}$Institute for Cosmic Ray Research, University of Tokyo, Tokyo, Japan\\
$^{23}$Deutsches Elektronen-Synchrotron DESY, Hamburg, Germany\\
$^{24}$Max-Planck-Institut f\"{u}r Physik, Munich, Germany\\
$^{25}$Department of Nuclear Engineering, University of California Berkeley, Berkeley, CA, USA\\
$^{26}$Fermi National Accelerator Laboratory, Batavia, IL, USA\\
$^{27}$Physics Department, University of Patras, Patras, Greece\\
$^{28}$ Present address: Laboratoire de l'Acc\'el\'erateur Lin\'eaire, Centre Scientifique d'Orsay, Batiment 200 - BP34, 91898 Orsay Cedex, France}
\contribID{vogel\_julia}

\desyproc{DESY-PROC-2012-04}
\acronym{Patras 2012} 
\doi  

\maketitle

\begin{abstract}
The International Axion Observatory (IAXO) is a next generation axion helioscope aiming at a sensitivity to the axion-photon coupling of a few $10^{-12}$~GeV$^{-1}$, i.e. $1$-$1.5$ orders of magnitude beyond sensitivities achieved by the currently most sensitive axion helioscope, the CERN Axion Solar Telescope (CAST). Crucial factors in improving the sensitivity for IAXO are the increase of the magnetic field volume together with the extensive use of x-ray focusing optics and low background detectors, innovations already successfully tested at CAST. Electron-coupled axions invoked to explain the white dwarf cooling, relic axions, and a large variety of more generic axion-like particles (ALPs) along with other novel excitations at the low-energy frontier of elementary particle physics could provide additional physics motivation for IAXO.

\end{abstract}
\section{Introduction}
Quantum Chromodynamics (QCD) is expected to lead to CP-violation in flavor conserving interactions, but up to now no experiment has been able to observe this effect. A possible solution to this strong CP-problem was suggested by R.~Peccei and H.~Quinn~\cite{Peccei:1977a} over $30$~years ago. They explained the apparent conservation of CP in strong interactions by introducing an additional global gauge symmetry. When this symmetry is spontaneously broken at a yet unknown breaking scale $f_{a}$, it gives rise to a Nambu-Goldstone boson as was pointed out independently by S.~Weinberg and F.~Wilczek~\cite{Weinberg:1978}. This neutral pseudo-scalar is commonly referred to as the axion. The concept has been generalized to further particles (axion-like particles (ALPs)) which may arise as Nambu-Goldstone bosons from the breaking of other global symmetries.\\
%
Axions, if they exist, could have been created in the very early universe, but they could also still be produced in cores of stars like our Sun nowadays. Several constraints from astrophysics and cosmology have been considered, narrowing the most likely range of the symmetry breaking scale $f_a$  to a window reaching from $10^{9}$~GeV up to $10^{12}$~GeV. Experiments have attempted to detect axions in and close to the remaining regions in parameter space. Most of them make use of the Primakoff effect~\cite{Primakoff:1951}, which allows for a conversion of axions into photons in the presence of a strong electromagnetic field~\cite{Sikivie:1984}. Helioscopes are employing precisely this effect~\cite{VanBibber:1989}, looking for axions produced in the solar core. The most sensitive existing helioscope is the CERN Axion Solar Telescope (CAST~\cite{Zioutas:2005}), which utilizes a $10$~m long superconducting LHC prototype dipole magnet providing a magnetic field of up to $9$~T. The experiment is able to point at the Sun for a total of about $3$~h per day during sunset and sunrise. At both ends of the magnet, x-ray detectors have been mounted~\cite{Kuster:2007} to search for photons from Primakoff conversion. 
\section{The IAXO Experiment}
The proposed next generation axion helioscope, dubbed the International AXion Observatory (IAXO), promises to be the most sensitive detector for solar axions ever built~\cite{Irastorza:2011}. For hadronic axion models, in which axions do not couple to electrons at tree level (the main focus of helioscopes), the anticipated sensitivity to the axion-to-photon coupling is expected to surpass the best current limits set by CAST by a factor of $10$-$20$, reaching coupling constants $g_{a\gamma}\sim \alpha/(2\pi f_a)$ of a few times $10^{-12}$~GeV$^{-1}$. Furthermore, a larger variety of models becomes testable, if axions also couple to electrons. With IAXO reaching its operational goals, it may probe QCD axions model regions with masses smaller than $10$-$20$ meV or provide evidence of exotic physics, like anomalous white dwarf (WD) cooling~\cite{Isern:2010}. Additionally, IAXO will be able to probe large unexplored regions of parameter space for ALPs and a variety of other (very) weakly interacting sub-eV particles~\cite{Jaeckel:2010ni}, for example hidden photons~\cite{Gninenko:2008pz}. It is almost guaranteed that any particle found by IAXO including the standard QCD axion is at the very least a subdominant component of dark matter, and, depending on the models, it could even account for all or most of it~\cite{Arias:2012az}.
The key features to improving on current helioscope experiments are reflected by a suitable figure of merit (FOM) as defined in Ref.~\cite{Irastorza:2011} using the dependence of the axion-photon coupling on experimental components of a helioscope, namely,
\begin{equation}
g^{4}_{a\gamma}\propto \underbrace{(BL)^{-2}A^{-1}}_{Magnet} \times \underbrace{b^{1/2}\epsilon^{-1}}_{Detectors} \times \underbrace{s^{1/2}\epsilon_{0}^{-1}}_{X-ray~optics} \times \underbrace{t^{-1/2}}_{Exposure}
\end{equation}
where $B$, $L$ and $A$ are the magnetic field strength, the length and cross-sectional area of the magnet, respectively. The sensitivity of an axion helioscope also depends on the detector background $b$ and efficiency $\epsilon$ as well as the focal spot size $s$ of the x-ray optics in place and the telescope efficiency $\epsilon_{0}$. The last contribution originates from the ability of the experiment to follow the Sun and is included in form of the exposure time $t$. With this in mind, the minimum requirements for a helioscope of significantly improved sensitivity are a powerful magnet of large volume and an appropriate x-ray sensor covering the exit of the magnet bore. Ideally, the magnet is equipped with a mechanical system enabling it to follow the Sun and thus increasing the exposure time. The IAXO experiment will take all these factors into consideration to maximally increase sensitivity as compared to state-of-the-art solar axion searches.
\subsection*{Experimental Setup and Expected Sensitivity}
The basic IAXO setup will resemble the conceptual layout of an enhanced axion helioscope: it is envisioned to consist of a toroidal magnet coupled via x-ray optics to detectors attached
to each of the magnet bores. 
For all experimental parameters improvements over current best values are anticipated (details can be found in Ref.~\cite{Irastorza:2011}) yielding to four IAXO scenarios that range from most conservative to most optimistic. While a detailed justification of the assumed parameters for these scenarios can be found in Ref.~\cite{Irastorza:2011}, this paper will briefly outline the most important considerations before introducing the four IAXO scenarios.\\
\indent{\bf Magnet:} The magnet parameters are the determining factors of a helioscope's FOM. The availability of a strong LHC test magnet that has been recycled to become part of CAST, is one of the main reasons for the success of this helioscope. In order to surpass the CAST sensitivity building a custom-designed magnet is essential. Improving on $B$ and $L$ of the CAST magnet is extremely challenging, but increasing the cross sectional area of the magnet is a promising approach, since the area $A_{\rm CAST}$ is only $3 \times 10^{-3}$~m$^{2}$. While a change in the magnet configuration is needed to achieve significantly larger cross sections, this approach is feasible with the design and construction of a new magnet. A toroidal configuration for the IAXO magnet is being studied~\cite{Shilon:2012}. Its total cross sectional area $A$ can be up to a few m$^{2}$, while the product $BL$ is kept close to levels achieved for CAST.\\
\indent{\bf X-Ray optics:} Although CAST has proven the concept, only one of the four magnet ports of the experiment is currently equipped with x-ray optics. The impact on sensitivity of using focusing power for the entire magnet cross-section $A$ is implicit in the FOM defined above (see Eq.~(1)). In part the improvement obtained by enlarging $A$ results from the fact that a correspondingly larger optic is coupled to the magnet. The optics challenge in this case is two-fold: not only must we optimize the optic term in the FOM (Eq.~(1)), but we must also consider the availability of cost-effective x-ray optics of the required size. IAXO's optics specifications can be met by a dedicated fabrication effort based on segmented glass substrate optics like the ones of HEFT or NuSTAR~\cite{Harrison:2000} and the design effort for this is currently ongoing.\\
\indent{\bf Detectors:} During the life of CAST its detectors have been constantly developed with the goal of lowering the detector background level. Latest generations of Micromegas detectors in CAST are able to achieve background levels as low as $5\times 10^{-6}$~counts keV$^{-1}$cm$^{-2}$s$^{-1}$. This value is already a factor $20$ better than the background levels measured during the first data-taking periods of CAST. It appears feasible to reduce this level to or even below $10^{-7}$~counts keV$^{-1}$cm$^{-2}$s$^{-1}$ .
\begin{figure}[bt!]
\centering
\includegraphics[width=0.87\textwidth]{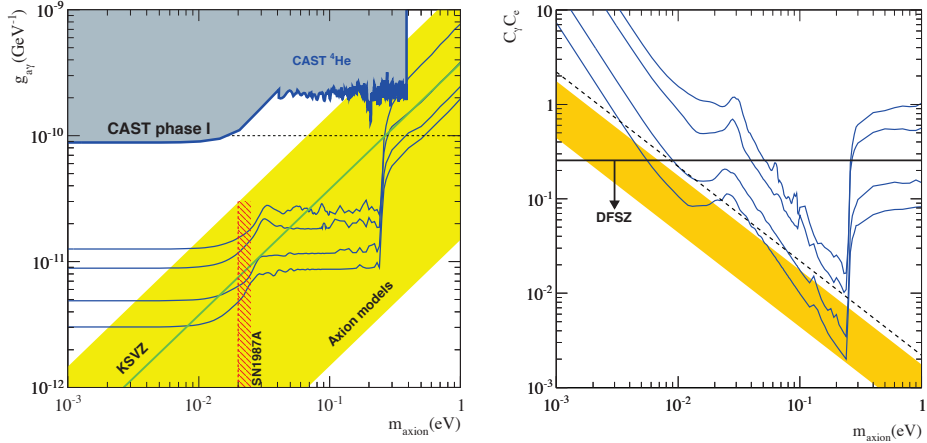}
\caption{Left: Parameter space for axions and ALPs. The thin blue lines indicate the expected IAXO sensitivity, ranging from conservative to best case (see Ref.~\cite{Irastorza:2011} for a details on the four scenarios). The CAST limit, some astrophysical limits, along with the range of standard QCD axion models (yellow band) are shown as well.  Right: The expected sensitivity of the same four IAXO scenarios in the parameter space of axions and ALPs if they also couple to electrons. For GUT models ($C_{\gamma}=0.75$) the bound on the electron coupling ($C_{e}$) from red giants is shown (dashed line) along with the region motivated by WD cooling (orange band). DFSZ models lie below the horizontal black line.}\label{Fig:IAXO}
\label{sec:figures}
\end{figure}
\\The accessible region of parameter space for IAXO would include a large fraction of realistic, previously unexplored QCD ranges of axion model space at the meV mass scale and above. Especially relevant is the possibility to directly test the axion hypothesis invoked to solve the anomalous cooling rate of white dwarfs~\cite{Isern:2010}, favoring electron-coupled axions with a mass of about $17$ meV. This region is out of reach of any other existing experimental technique. The computed sensitivities for different IAXO scenarios (conservative, realistic, optimistic and best case) are represented in Fig.~\ref{Fig:IAXO} by a family of blue lines, both for hadronic axions (left) and non-hadronic ones (right). For each scenario, two data taking campaigns are included: a three year long period of data acquisition performed with vacuum in the magnetic field region (analogous to CAST Phase I), and a second three year campaign with varying amounts of $^{4}$He gas inside the magnet bore (analogous to the first part of CAST's Phase II).\\
In summary, IAXO sensitivity lines are expected to go well beyond current CAST limits for hadronic axions. Additionally, they are progressively advancing into the decade of $10^{-11}$-$10^{-12}$~GeV$^{-1}$. With this the experiment is sensitive to QCD axion in realistic models at the meV mass scale and could exclude a large portion of such models above this range. For non-hadronic axions, the IAXO sensitivity lines start exploring the DFSZ model region and the experiment could provide constraints as good as or even better than those obtained from red giants~\cite{Raffelt:1995}. Most importantly, the two more optimistic scenarios for IAXO start probing the region of parameter space highlighted by the cooling of white dwarfs. Additionally, if the x-ray detection technology can achieve a low-energy threshold of about $0.1$~keV, IAXO could also test theories that invoke solar processes to create chameleons~\cite{Brax:2010}. More speculative, but of potentially significant scientific gain, is the construction of IAXO in such a way that it could accommodate microwave cavities to operate inside this powerful magnet allowing for a simultaneous search of solar and relic axions, present in the Galactic halo. \vspace{1.6mm}
\\ 
{\bf Acknowledgments} This article has been authored by Lawrence Livermore National Security, LLC under Contract No. DE-AC52-07NA27344 with the U.S. Department of Energy.  Accordingly, the United States Government retains and the publisher, by accepting the article for publication, acknowledges that the United States Government retains a non-exclusive, paid-up, irrevocable, world-wide license to publish or reproduce the published form of this article or allow others to do so, for United States Government purposes.


\begin{footnotesize}

\end{footnotesize}


\end{document}